\documentclass[12pt]{article}
\usepackage{subfigure}
\usepackage{amssymb,amsmath}
\usepackage{graphicx}
\usepackage{float}
\usepackage{color}
\usepackage{array,multirow}
\usepackage[font=footnotesize]{caption}
\usepackage[colorlinks=true
,urlcolor=blue
,citecolor=blue
,linkcolor=blue
,pagecolor=blue
,linktocpage=true
,pdfproducer=medialab
]{hyperref}
\usepackage[a4paper,width=16cm]{geometry}
\usepackage{cite}
\usepackage{placeins}
\usepackage[utf8]{inputenc}
\usepackage{cancel}
\usepackage{arydshln}
\makeatletter \renewcommand{\@dotsep}{10000} \makeatother
\usepackage{indentfirst}
\usepackage{rccol}
\usepackage{pifont}

\def\beq{\begin{equation}}
\def\eeq{\end{equation}}
\def\ba{\begin{array}}       
\def\ea{\end{array}}
\def\bea{\begin{eqnarray}}   
\def\eea{\end{eqnarray}}
\usepackage[nodisplayskipstretch]{setspace}

\begin{document}

\begin{titlepage}
\pagestyle{empty}

\vspace*{0.2in}
\begin{center}
{\Large \bf Higgs Information and NMSSM at the Large Hadron Collider}\\[0.25cm]

\vspace{1cm}

{\bf {Surabhi Gupta\footnote{E-mail: sgupta2@myamu.ac.in} and Sudhir Kumar Gupta\footnote{E-mail: 
sudhir.ph@amu.ac.in} }}

\vspace{2pt}
	\begin{flushleft}
		{\em Department of Physics, Aligarh Muslim University, Aligarh, UP--202002, India} 
	\end{flushleft}
	
	\vspace{10pt}

\begin{abstract}
Information theory has proven to be a worthwhile tool for investigating the implications of the Higgs sector in the Next-to-minimal supersymmetric Standard Model (NMSSM) using Higgs information at the Large Hadron Collider assessed through the entropy constructed by means of the branching ratios of decay channels of the Higgs boson. The present article focuses on the parameter space of supersymmetric extension with an extra term of gauge singlet in light of various experimental constraints. Our findings show the most preferred values of $m_0$, $m_{1/2}$, $ A_0$, $ tan\beta$,  $\lambda$, $\mu_{eff}$, neutralino LSP $ m_{\tilde\chi^{0}_{1}}$, lightest chargino $ m_{\tilde\chi^{\pm}_{1}}$, singlino $ m_{\tilde\chi^{0}_{5}}$, and gluino $ m_{\tilde g}$ to be around 1.93 TeV, 1.78 TeV, $-$3.62 TeV, 27.5, 0.012, 665.7 GeV, 0.74 TeV, 0.79 TeV, 11.24 TeV, and 3.70 TeV, respectively, that is compatible with the relic density of dark matter. 

\end{abstract}
\end{center}
\end{titlepage}

\section{Introduction}
\label{sec:intro}
The discovery of Higgs Boson in 2012 by the ATLAS and CMS detector at the Large Hadron Collider (LHC)~\cite{ATLAS:2012yve, CMS:2012qbp, Aad:2015zhl} could provide mass to the elementary particles of the Standard Model (SM)~\cite{Djouadi:2005gi}. Although the SM lacks a mechanism that could stabilise the Higgs mass against the radiative correction and explain the interesting natural phenomena such as grand unification, baryogenesis, neutrino oscillations, and the existence of dark matter. Supersymmetry (SUSY)~\cite{Martin:1997ns, Tata:1997uf, Drees:1996ca, Aitchison:2005cf, Djouadi:2005,Fayet:2015sra, Cane:2019ac, Allanchach:2019wrx} is one of the potential extensions of the SM and can provide the solutions to the aforesaid problems such as providing stability to the Higgs boson mass against the radiative corrections, giving the viable dark matter candidate, and explicating the grand unification successfully. One of the models of the extended Higgs sector such as the Minimal Super­symmetric Standard Model (MSSM)~\cite{Aitchison:2005cf, Djouadi:2005, Nilles:1983ge, Haber:1984rc, Dawson:1997tz, Heinemeyer:1998np, Draper:2016pys} can be able to acquire the 125 GeV Higgs boson that contains the two Higgs doublets contributing to five Higgs states. In the MSSM, the Higgs boson of 125 GeV would acquire through the radiative corrections, particularly the stop loops. The $\mu$ parameter is introduced to provide a mass to the complex two SU(2) Higgs doublets $H_u$ and $H_d$ present in the Lagrangian of MSSM, however, requires to be of the order of electroweak scale. 

Therefore, through the extension to MSSM such as the Next-to-minimal supersymmetric Standard Model (NMSSM)~\cite{Ellwanger:2009dp,Maniatis:2009re} contains an extra term of Higgs singlet field which could possibly solve the $\mu$ problem~\cite{Kim:1983dt} by generating the $\mu$ term dynamically in the SUSY breaking scale. Furthermore, the Higgs sector of NMSSM consists of seven Higgs states including three CP-even Higgses, two CP-odd Higgses, and two charged Higgses through the two Higgs doublets $H_u$ and $H_d$ with one added Higgs singlet field $S$. This could also further increase the upper limit of Higgs mass at the tree level along with the spectrum enhanced with the two scalars Higgs bosons and one fermionic neutralino. The NMSSM is also successful to address some of the problems such as the Grand unification and provides the lightest supersymmetric particle (LSP) to be the dark matter candidate~\cite{Maniatis:2009re}.
Moreover, NMSSM has been extensively studied at the LHC~\cite{Ellwanger:2005uu, Ellwanger:2011sk, Heng:2023xfb, Beskidt:2017dil, Barman:2020vzm, Biekotter:2021qbc,Tang:2022pxh}.

In the current study, we will focus on the NMSSM scenario using the information entropy of the lighter CP-even Higgs boson assuming it to be the LHC-observed Higgs boson.
Information entropy has been investigated in particle physics to accomplish remarkable results. In the study~\cite{dEnterria:2012eip}, the SM Higgs mass has been evaluated through maximisation of the product of all its branching ratios relating to the maximum possible decays which is well in agreement with the LHC observed Higgs mass. Moreover, in Ref.~\cite{Alves:2014ksa}, the mass of the SM Higgs boson was observed through the maximisation of the information entropy constructed using the branching ratios of the SM Higgs boson.
Further Maximum Entropy Principle (MEP) has also been applied in the study of observing new modes of decay at the LHC related to the Higgs boson~\cite{Alves:2020cmr} and particles~\cite{Millan:2018fme,Llanes-Estrada:2017clj}, studying the axion mass~\cite{Alves:2017ljt}, and successfully investigated the SUSY models~\cite{Gupta:2020whs,Gupta:2022psc,Gupta:2022mjt}. In this study, information entropy is employed to examine the NMSSM scenario in more detail that could be built using the branching ratios of the Higgs boson predicting a Higgs mass at its maximum value. In addition, Higgs entropy could be explored as a tool for predicting sparticle masses under various experimental constraints. 

The paper is organised as follows. In Section II, we explore the NMSSM scenario related to Higgs boson. In Section III, we investigate the information entropy in the context of CP-even lightest Higgs boson. In Section IV, we discuss the NMSSM scenario presented in terms of information theory. In Section V, we conclude our results.

\section{Higgs bosons in the NMSSM}
We have taken into account NMSSM or (M+1)SSM specified as the MSSM supplemented with an extra gauge singlet chiral superfield $\widehat{S}$. The scale invariant renormalisable superpotential for the NMSSM~\cite{Ellwanger:2009dp,Maniatis:2009re} containing the renormalisable couplings is characterized by
\beq
 \mathcal{W} = \mathcal{W}_{\text{MSSM}}^{\cancel{\mu}} + \lambda \widehat S \widehat H_u \cdot \widehat H_d + \frac{1}{3} \kappa \widehat S^3,
 \label{2.1e}
\eeq 
where the first term, $\mathcal{W}_{\text{MSSM}}^{\cancel{\mu}}$, defines the Yukawa couplings the same as in the MSSM superpotential, $\mu$ term in the MSSM is substituted by the above second term that is produced dynamically through a vacuum expectation value (VEV) $s$ of singlet superfield $\widehat S$ forming a supersymmetric mass term $\mu_{eff} = \lambda s$, in which $\lambda$ represents a coupling among $\widehat S$, $\widehat H_u$ (up type Higgs superfield) and $\widehat H_d$ (down type Higgs superfield) and the third term denotes a cubic self-coupling in $\widehat S$ in which the dimensionless coupling parameter $\kappa$ is added in order to have $\mathbb{Z}_3$ invariance. Here, the $\mu_{eff}$ is of the order of the SUSY breaking scale, that fixes the $\mu$ problem of the MSSM. The soft-SUSY breaking masses and couplings included in the Lagrangian of the NMSSM are given below~\cite{Ellwanger:2009dp,Maniatis:2009re}
\begin{eqnarray}
\nonumber -\mathcal{L}^{\text{\tiny NMSSM}}_{\text{\tiny soft}}&=&\dfrac{1}{2}\left(M_1\tilde{\lambda}_1\tilde{\lambda}_1+M_2\tilde{\lambda}_2^i\tilde{\lambda}_2^i+M_3\tilde{\lambda}_3^a\tilde{\lambda}_3^a+h.c.\right)\\
\nonumber &&+ m_{\tilde{Q}}^2|\tilde{Q}|^2+m_{U}^2|\tilde{U}_{R}|^2+ m_{\tilde{D}}^2|\tilde{D}_{R}|^2+ m_{\tilde{L}}^2|\tilde{L}|^2+m_{\tilde{E}}^2|\tilde{E}_{R}|^2\\
\nonumber&&+m^2_{H_u}\left|H_u\right|^2+ m^2_{H_d}\left|H_d\right|^2+m^2_{S}\left|S\right|^2\\
\nonumber&&+ h_uA_u\tilde{U}_{R}^c\tilde{Q}\cdot H_u -h_dA_d\tilde{D}_{R}^c\tilde{Q}\cdot H_d\\
&&-h_eA_e\tilde{E}_{R}^c\tilde{L}\cdot H_d+\lambda A_{\lambda}\ S H_u\cdot H_d+\dfrac{1}{3}\kappa A_{\kappa} S^3 +h.c. .\label{2.2e}
\end{eqnarray} 	
The first line describes the gaugino mass terms $M_i$ representing its respective couplings $\tilde{\lambda}_1$, $\tilde{\lambda}_2^i$ $(i=1,2,3)$, and $\tilde{\lambda}_3^a$ $(a=1,...,8)$ corresponds to the $U(1)_Y$, $SU(2)$, and $SU(3)$ gauginos, respectively. The second line contains squarks and sleptons mass squared terms of $3\times 3$ Hermitian matrices as $m_{\tilde{Q}}^2$, $m_{U}^2$, $m_{\tilde{D}}^2$, $m_{\tilde{L}}^2$, and $m_{\tilde{E}}^2$ while the third line represents the terms of the Higgs mass squared, $m^2_{H_u}$ and $m^2_{H_d}$, and singlet mass squared, $m^2_{S}$. Further ${h_{u,d,e}}$, $\lambda$, and $\kappa$ correspond to dimensionless Yukawa couplings whereas $A$ parameters represent a mass dimension trilinear scalar interaction. Trilinear interaction terms ($A_{\lambda}$ and $A_{\kappa}$) and soft-SUSY breaking mass term related to the singlet have been included to the NMSSM soft-SUSY breaking Lagrangian besides the MSSM one. In this Lagrangian, the $SU(2)$~doublets correspond to  
\beq\label{2.3e}
\widehat{Q} = \left(\ba{c} \widehat{U}_L \\ \widehat{D}_L
\ea\right) , \
\widehat{L} = \left(\ba{c} \widehat{\nu}_{L} \\ \widehat{E}_L
\ea\right) , \
\widehat{H}_u = \left(\ba{c} \widehat{H}_u^+ \\ \widehat{H}_u^0
\ea\right) , \
\widehat{H}_d = \left(\ba{c} \widehat{H}_d^0 \\ \widehat{H}_d^-
\ea\right) ,
\eeq
and the products of these $SU(2)$ doublets can be specified as, 
\beq\label{2.4e}
\widehat{H}_u \cdot \widehat{H}_d = \widehat{H}_u^+ \widehat{H}_d^- 
- \widehat{H}_u^0 \widehat{H}_d^0\ .
\eeq

In addition to the MSSM particles, NMSSM consists of $P_R=-1$ Weyl fermion i.e. singlino, $P_R=+1$ a real and a pseudo scalar. The neutralino sector has also been impacted by the newly added terms to the NMSSM where the four MSSM neutralinos combine with the singlino $\tilde S$, odd $R$-parity, to produce five neutralinos. The neutralino mass matrix is $5\times5$ instead of $4\times4$ in the MSSM under the gauge-eigenstate basis $\psi^0 = (\tilde B, \tilde W^0, \tilde H_d^0, \tilde H_u^0, \tilde S)$ which could be given as
\beq
{\bf M}_{\tilde N} \,=\, \begin{pmatrix}
  M_1 & 0 & -g' v_d/\sqrt{2} & g' v_u/\sqrt{2} & 0\cr
  0 & M_2 & g v_d/\sqrt{2} & -g v_u/\sqrt{2} & 0\cr
  -g' v_d/\sqrt{2} & g v_d/\sqrt{2} & 0 & -\lambda s & -\lambda v_u\cr
  g' v_u/\sqrt{2} & -g v_u/\sqrt{2}& -\lambda s & 0 & -\lambda v_d\cr 
  0 & 0 & -\lambda v_u & -\lambda v_d & 2 \kappa s
  \end{pmatrix}
\label{2.5e}
\eeq
After diagonalisation of the above matrix, the mass eigenstates are $\tilde{\chi}_1^0, \dots, \tilde{\chi}_5^0$, from which the lightest one is considered as the LSP as well as the good candidate for the dark matter and would be written as 
\begin{equation}
\tilde{\chi}_1^0=N_{11}\tilde{B}+N_{12}\tilde{W}^0+ N_{13}\tilde{H}_d^0+N_{14}\tilde{H}_u^0+N_{15}\tilde{S},\\
\end{equation}
where $N_{1j}$ corresponds to matrix elements of the diagonalization matrix and sum of matrix elements, $\sum |N_{1j}|^2 =1$.
The masses of gaugino are achieved by means of $M_1$, $M_2$, and $M_3$ soft SUSY-breaking mass parameters whereas the masses of higgsino are acquired with the help of the $\mu$ parameter. The NMSSM contains an extra fermion term singlino which is the fifth neutralino. 
Through the mixing of Higgsinos and singlino (fermionic superpartners correspond to Higgs fields) with the gauginos (fermionic superpartners correspond to gauge bosons) producing five neutral fermions termed neutralinos while two charged fermions termed charginos. 
In the NMSSM, the Higgs fields include the two Higgs doublets along with Higgs singlet $S$. The impact of substituting a $\mu$ term with the dynamic field $S$, the correction to the Higgs mass at tree-level, through an extra F-term proportional to $\lambda$ in the scalar potential, has been increased by an amount   
\beq
\Delta(m_{h}^2) \leq \lambda^2 v^2 \sin^2 (2 \beta).
\label{2.6e}
\eeq
Due to an additional singlet superfield contribution in the Higgs sector, the number of Higgs bosons has raised to seven, namely three neutral CP-even, two neutral CP-odd, and two charged Higgs bosons.

The Higgs potential would be achieved with SUSY gauge interactions, soft SUSY breaking, and $F$-terms as follows~\cite{Ellwanger:2009dp,Maniatis:2009re}
\bea
V_\mathrm{Higgs} & = & \left|\lambda \left(H_u^+ H_d^- - H_u^0
H_d^0\right) + \kappa S^2 + \mu' S +\xi_F\right|^2 +\left(m_{H_u}^2 + \left|\mu + \lambda S\right|^2\right) 
\left(\left|H_u^0\right|^2 + \left|H_u^+\right|^2\right)\nonumber \\
&&+\left(m_{H_d}^2 + \left|\mu + \lambda S\right|^2\right) 
\left(\left|H_d^0\right|^2 + \left|H_d^-\right|^2\right) 
+\frac{g_1^2+g_2^2}{8}\left(\left|H_u^0\right|^2 + 
\left|H_u^+\right|^2 - \left|H_d^0\right|^2 -
\left|H_d^-\right|^2\right)^2 \nonumber  \\
&& +\frac{g_2^2}{2}\left|H_u^+ H_d^{0*} + H_u^0 H_d^{-*}\right|^2 +m_{S}^2 |S|^2
+\big( \lambda A_\lambda \left(H_u^+ H_d^- - H_u^0 H_d^0\right) S + 
\frac{1}{3} \kappa A_\kappa\, S^3 \nonumber \\
&&+ m_3^2 \left(H_u^+ H_d^- - H_u^0 H_d^0\right) 
 +\frac{1}{2} m_{S}'^2\, S^2 + \xi_S\, S + \mathrm{h.c.}\big)
\label{2.7e}
\eea
with $g_1$ and $g_2$ are associated with $U(1)_Y$ and $SU(2)$ gauge couplings,
respectively. Neutral three CP-even and two CP-odd Higgs bosons are produced as a result of the mixture of the neutral components of the two Higgs doublets and singlet.  
The neutral Higgs doublets and singlet in terms of components can be attained through the expansion of the Higgs potential in terms of $v_u$, $v_d$, and s which would be written as
\beq\label{2.8e}
H_u^0 = v_u + \frac{H_{uR} + iH_{uI}}{\sqrt{2}} , \qquad
H_d^0 = v_d + \frac{H_{dR} + iH_{dI}}{\sqrt{2}} , \qquad
S = s + \frac{S_R + iS_I}{\sqrt{2}}\ .
\eeq
where index $R$ and $I$ signify for CP-even and CP-odd states, while $v_u$, $v_d$, and $s$ denote the real neutral VEVs. The mixing of real parts of Singlet Higgs field $S$ and Higgs doublet $H_u^0$ and $H_d^0$ form CP-even Higgs bosons whereas the mixing of imaginary parts of these Higgs fields produce CP-odd Higgs bosons. The lower mass limit on lightest chargino from LEP searches puts a lower limit of 100 GeV on $\mu$~\cite{Ellwanger:2009dp}.

\section{Higgs Information and the NMSSM} 
The information theory~\cite{jaynes:1957,thomas:2006} is employed to measure the uncertainty of a state by using Shannon's~\cite{shannon} entropy or information entropy. In other words, it is a lack of information, a disorder resulting in an increasing amount of entropy or randomness. It is fundamentally a theory of probabilities in which each probability stands for the ignorance of an event where the probability distribution includes information on each event. As such the amount of information or entropy is considered to be the expected value for the event's random variable where the information is regarded as negative with respect to the logarithm of the probability distribution. The probabilities range from zero to one, and the sum of all probabilities equals one since each event is mutually independent and exhaustive. Probabilities are associated with uncertainties in the system. Since probable events are linked to less information or less uncertainty, more information is provided by rarer events. If the outcome of an event is already known, it produces zero information or zero entropy. Maximum entropy is an equilibrium state producing maximum information or maximum uncertainty. Using the maximum entropy principle (MEP), one can estimate the best value of the variable of the probability distribution constructed using Shannon’s entropy. The MEP has been discussed in detail for the analysis of the Higgs boson concerning models such as CMSSM~\cite{Gupta:2020whs}, NMFV with CMSSM~\cite{Gupta:2022psc}, and Split-SUSY~\cite{Gupta:2022mjt}. 

For the current work, we have taken into account Shannon's entropy estimating through branching ratios of decay channels of the Higgs boson, namely $h\rightarrow\gamma\gamma$, $h\rightarrow \gamma Z$, $h\rightarrow Z Z^*$, $h\rightarrow W W^*$, $h\rightarrow gg$, $h\rightarrow f\bar{f}$ with $f\in \{u, d, c, s, b, e, \mu, \tau \}$, $h\rightarrow A A$. We examine an ensemble consisting of $\cal N$-independent Higgs bosons concerning the information theory that might decay into aforementioned decays modes of branching ratio $Br_j (m_h)$ contributing to its respective probability $p_{_j} (m_h)$,  
$p_{_j} (m_h) \equiv Br_j (m_h) =  \frac{\Gamma_j (m_h)}{\Gamma (m_h)}$, where $\Gamma_j (m_h)$ denotes partial decay width corresponding to $j^{th}$ decay channel,
$\Gamma (m_h) = \sum_{j = 1}^{n_j}\Gamma_{j}(m_h)$ considers the total decay width of Higgs that decays to all decay channels, and $n_j$ refers to the total number of decay channels of the Higgs boson. The probability of each Higgs boson of the concerned ensemble decays to its possible decay channels in the form of the multinomial distribution, which could be written as~\cite{Alves:2014ksa}
\beq
\label{3.1e}
{\cal P}_{\{m{_j}\}}(m_h) = \frac{{\cal N}!}{m_1!...m_{n_j}!}\prod_{j = 1}^{n_j}{(p_{j} (m_h))}^{m_j}, 
\eeq
where $\sum^{n_j}_{j = 1} {Br}_{j} = 1 $, $ \sum_{j = 1}^{n_j}m_{j} = {\cal N}$, $m_{j}$ denotes the number of Higgs bosons decay in $j^{th}$ mode.
The Shannon entropy would be written as~\cite{Alves:2014ksa} 
\beq
\label{3.2e}
S (m_h) = - \sum^{\cal N}_{\lbrace m_j \rbrace} {\cal P}_{\{m{_j}\}}(m_h) \ln {\cal P}_{\{m{_j}\}}(m_h).
\eeq
Hence an asymptotic expansion of the aforementioned entropy would be specified as~\cite{Alves:2014ksa}
\beq
\label{3.3e}
S (m_h) \simeq \frac{1}{2}\ln\left(\left(2\pi {\cal N} e\right)^{n_j -  1} \prod_{j = 1}^{n_j}{p_{j} (m_h)}\right) + \frac{1}{12 {\cal N}}\left( 3 n_j - 2 - \sum_{j = 1}^{n_j}{(p_{j} (m_h))}^{-1}\right)+ {\cal O}\left({\cal N}^{-2}\right).
\eeq

 \begin{table}[h]
 \begin{centering}
    \tabcolsep 0.4pt
    \small
    \begin{tabular}{cccc}
    \hline   
    \hline
    {Constraint}& {Observable}  & {Experimental~Value}& \qquad{Source}\\
       \hline 
LEP&$ m_{\tilde\chi^{0}_{1,2,3,4}}$ & $>$ 0.5 $m_Z$& \cite{Workman:2022ynf}\\
&$ m_{\tilde\chi^{\pm}_{1,2}}$& $>$ 103.5 GeV& \cite{Workman:2022ynf}\\
& $m_h$ &$>$ 114.4 GeV &\cite{ALEPH:2006tnd}\\
\hline
PO & $BR(B^0_s \to \mu^+\mu^-)$& $(3.0\pm0.4)\times10^{-9}$&  \cite{Zyla:2020zbs}\\
\hline
DM & $ \Omega_{\chi}h^{2}$ & $0.1197\pm0.0022$& \cite{Planck:2015fie}\\
       \hline
    \end{tabular}
     \caption{\sf{Experimental measurements of the different observables employed in this study.}}
      \label{tab:table1}
   \end{centering}
\end{table}


\section{Numerical Analysis}
In this section, we conduct extensive random scan sampling in the parameter space of NMSSM within the context of information theory. Information theory considers a tool for assessing the mass of the Higgs boson by means of the branching ratios of its decays. This can further be used to successfully estimate the masses of other sparticles. We have performed a detailed random scan of free parameters as discussed in the following limits 
\begin{itemize}
\item $m_0  \in [100, 3000] $ GeV,
\item $m_{1/2} \in [100, 4000] $ GeV,
\item $A_0   \in [-3000, 0]$ GeV,
\item $\tan{\beta} \in [1,30]$,
\item $\lambda \in [0.01,0.3]$,
\item $\kappa \in [0.05,0.3]$,
\item $\mu_{\text{eff}} \in [100, 3000]$ GeV, 
\end{itemize}
where $m_0$ and $m_{1/2}$ represent common scalar and gaugino masses at GUT scale, $A_0$ denotes common trilinear coupling at GUT scale, $\mu_{\text{eff}} \equiv \lambda s$ an effective Higgsino mass parameter at SUSY scale with $\lambda$ depicting the coupling of the Higgs doublets with the Higgs singlet at SUSY scale and $s$ denotes the VEV of Higgs singlet,  $\kappa$ denotes the dimensionless coupling parameter, and $\tan{\beta}$ indicates the ratio of VEVs of up- and down-type Higgs doublets ($v_u/v_d$). We assume $A_{\lambda}$ and $A_{\kappa}$ equal to zero.

Consequently, the presence of $\mu_{\text{eff}}$ in the order of electroweak scale solves the $\mu$-problem. Considering above-mentioned six free parameters, we use {\tt NMSSMTools (v6.6.0)}\cite{Ellwanger:2004xm, Ellwanger:2005dv,Belanger:2013oya} to generate the SUSY spectrum in which {\tt MicrOMEGAs} would also be used to measure the observables such as $ BR(B^0_s \to \mu^+\mu^-)$, $\Omega_{\chi}h^2$, spin-independent, and spin-dependent cross sections, the mass of Higgs boson and its branching ratios (involve exotic decays).

\begin{figure}[h]
\begin{centering}
\includegraphics[angle=0,width=0.5\linewidth,height=15em]{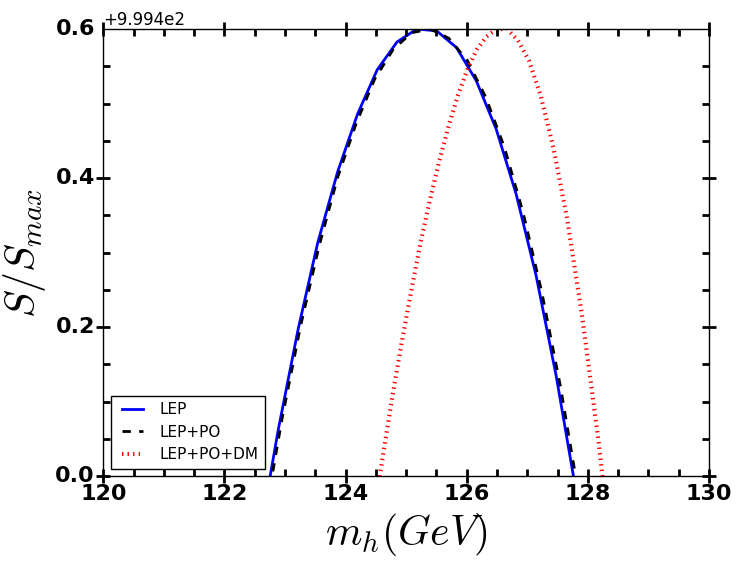}
\hspace{-0.5em}
\includegraphics[angle=0,width=0.45\linewidth,height=15.2em]{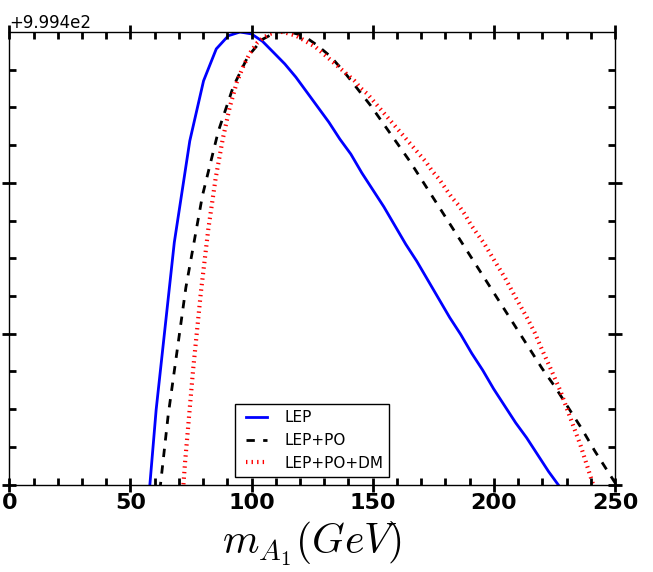}
\caption{\sf{Entropy vs lightest CP-even Higgs boson (left) and lightest CP-odd Higgs boson (right) corresponding to various constraints. The blue solid line includes the constraints of LEP data on Higgs boson, neutralino and chargino masses, the black dashed line includes the constraints of LEP data and B-Physics, and the red dotted line taking into account the constraints of LEP data, B-Physics, and the relic density of dark matter.}}
\label{fig1}
\end{centering}
\end{figure}
 
\begin{figure}[h]
\begin{centering}
\includegraphics[angle=0,width=0.33\linewidth,height=15em]{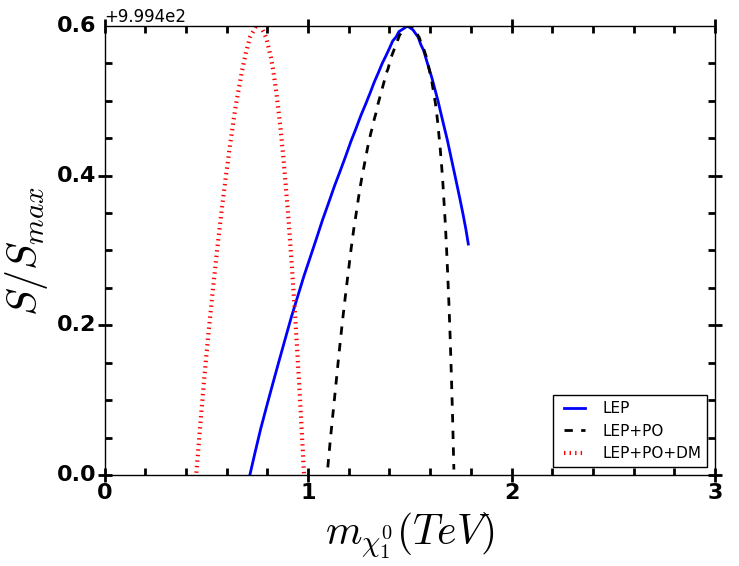}
\includegraphics[angle=0,width=0.327\linewidth,height=15em]{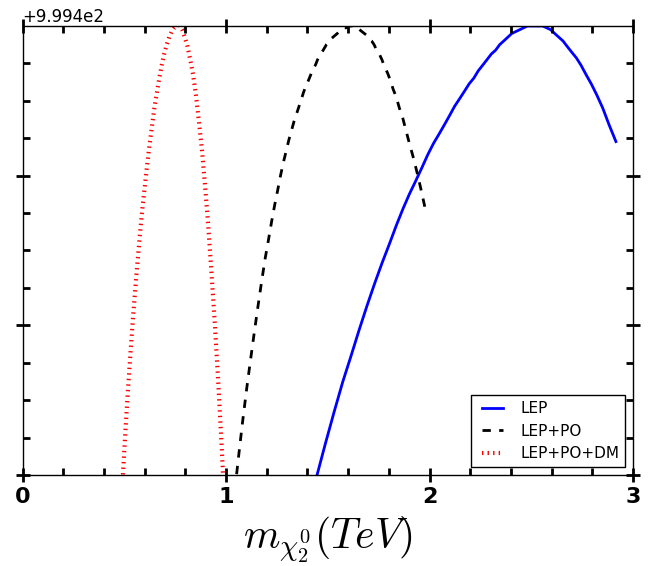}
\includegraphics[angle=0,width=0.327\linewidth,height=15em]{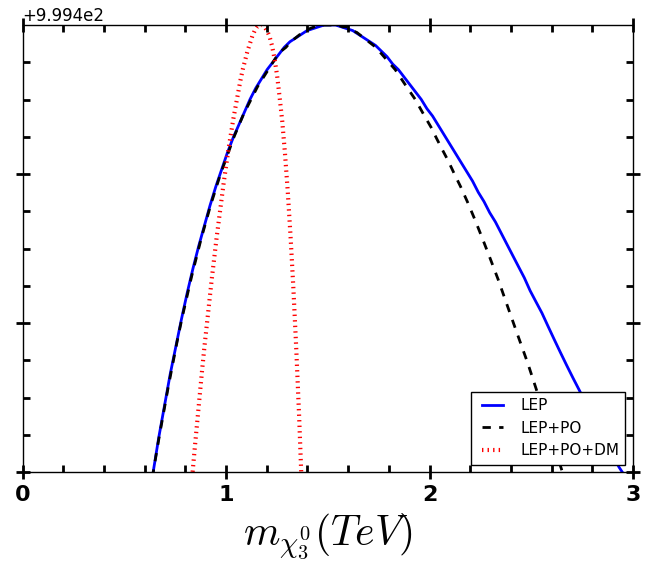}
\includegraphics[angle=0,width=0.33\linewidth,height=15.2em]{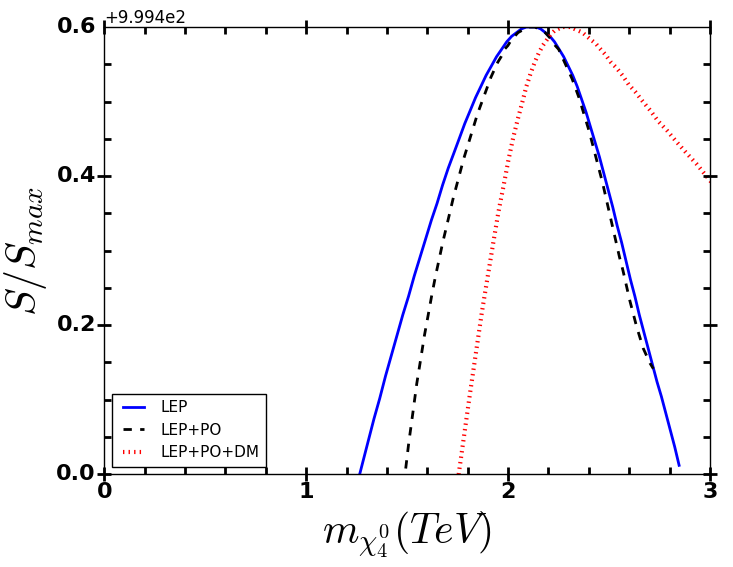}
\includegraphics[angle=0,width=0.327\linewidth,height=15.2em]{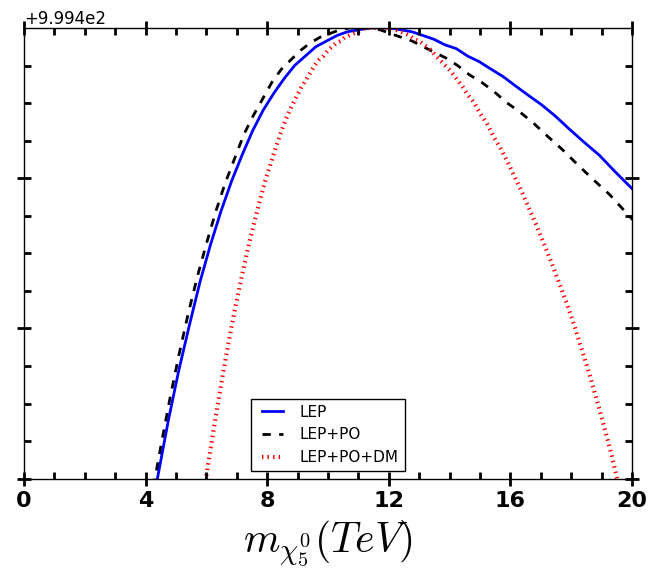}
\includegraphics[angle=0,width=0.327\linewidth,height=15em]{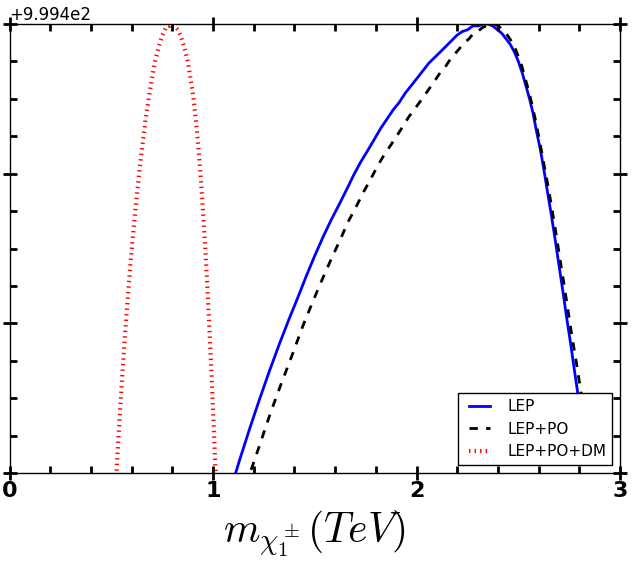}
\begin{center}
{
\includegraphics[angle=0,width=0.339\linewidth,height=15em]{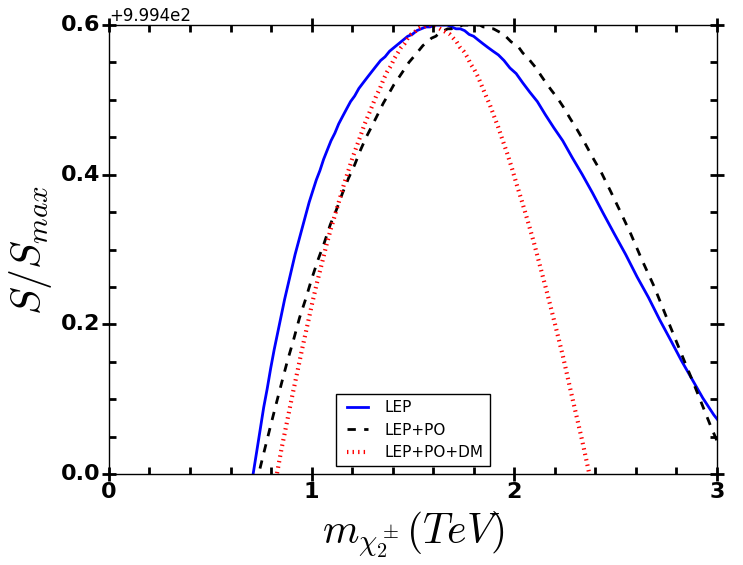}
\includegraphics[angle=0,width=0.327\linewidth,height=15em]{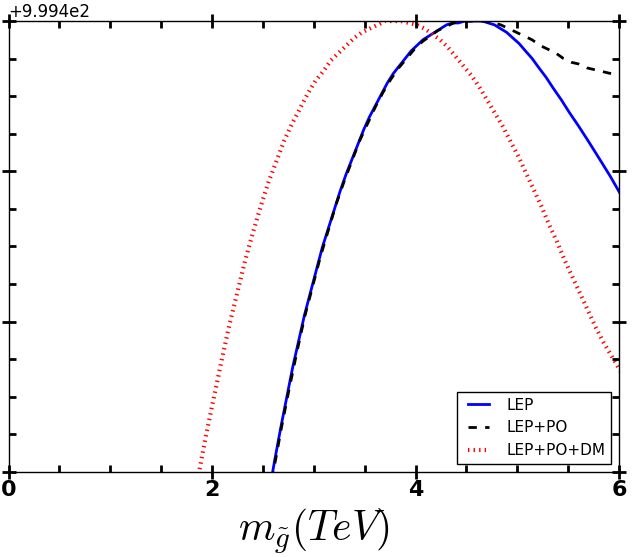}}
\end{center}
\caption{\sf{Entropy vs masses of gauginos corresponding to various constraints. The color scheme is same as the Figure~\ref{fig1}.}}
\label{fig2}
\end{centering}
\end{figure}

\begin{figure}[h]
\begin{centering}
\includegraphics[angle=0,width=0.33\linewidth,height=15em]{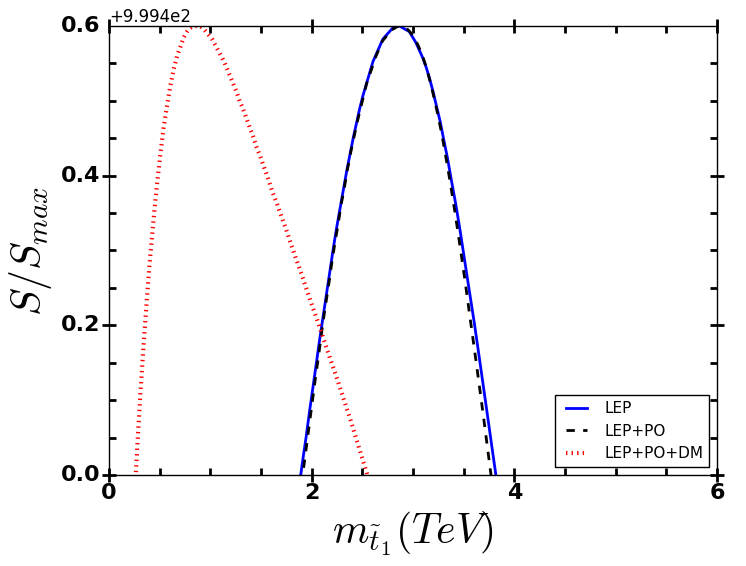}
\hspace{-0.9em}
\includegraphics[angle=0,width=0.327\linewidth,height=15em]{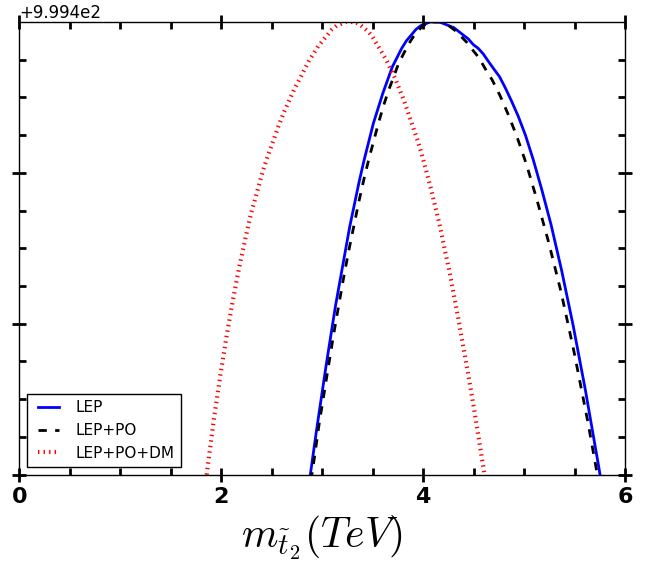}
\hspace{-0.9em}
\includegraphics[angle=0,width=0.327\linewidth,height=15em]{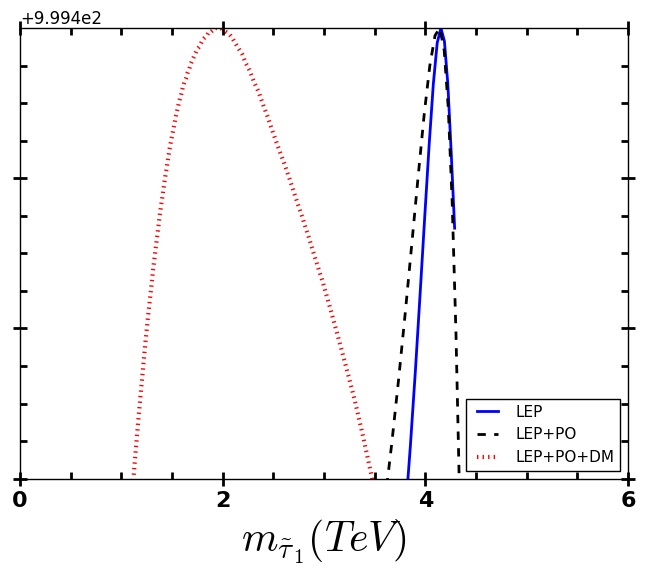}
\includegraphics[angle=0,width=0.33\linewidth,height=15.2em]{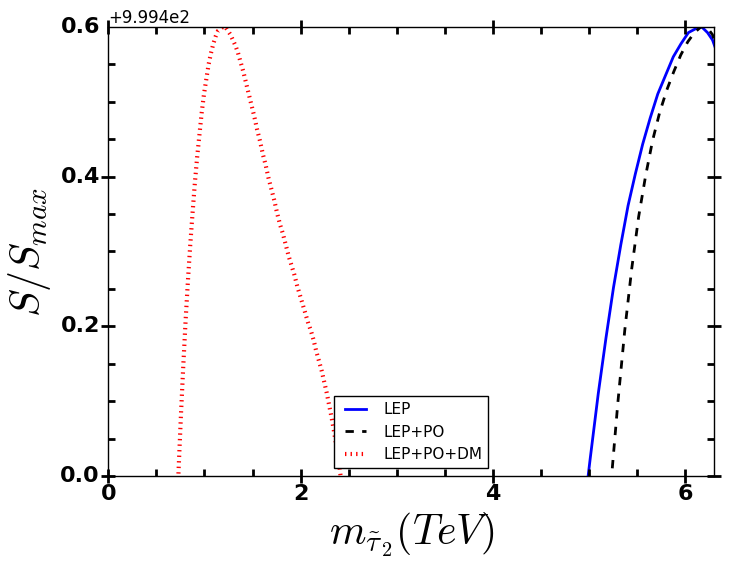}
\hspace{-0.9em}
\includegraphics[angle=0,width=0.327\linewidth,height=15em]{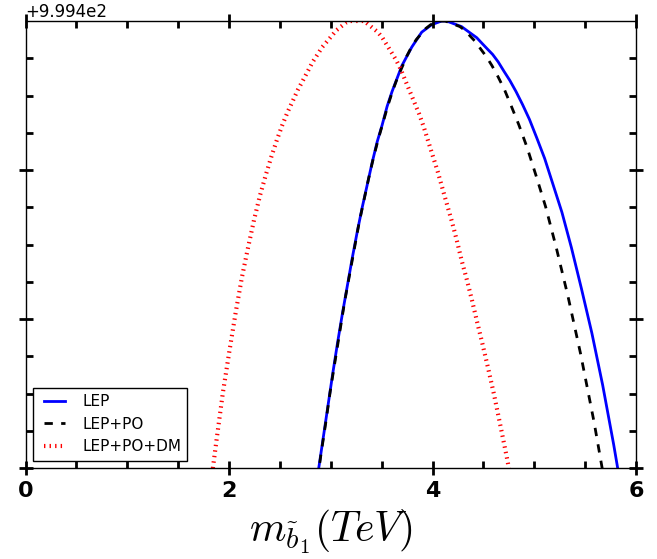}
\hspace{-0.9em}
\includegraphics[angle=0,width=0.327\linewidth,height=15em]{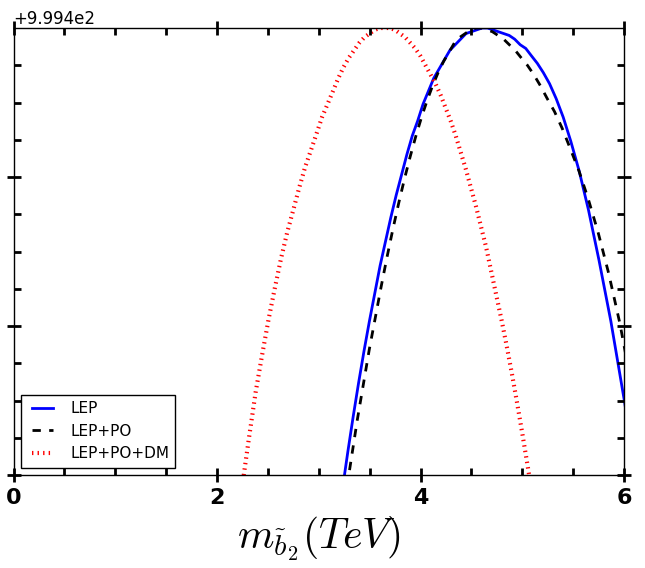}
\caption{\sf{Entropy vs masses of squarks corresponding to various constraints. The color scheme is same as the Figure~\ref{fig1}.}}
\label{fig3}
\end{centering}
\end{figure}

\begin{figure}[h]
\begin{centering}
\includegraphics[angle=0,width=0.327\linewidth,height=15em]{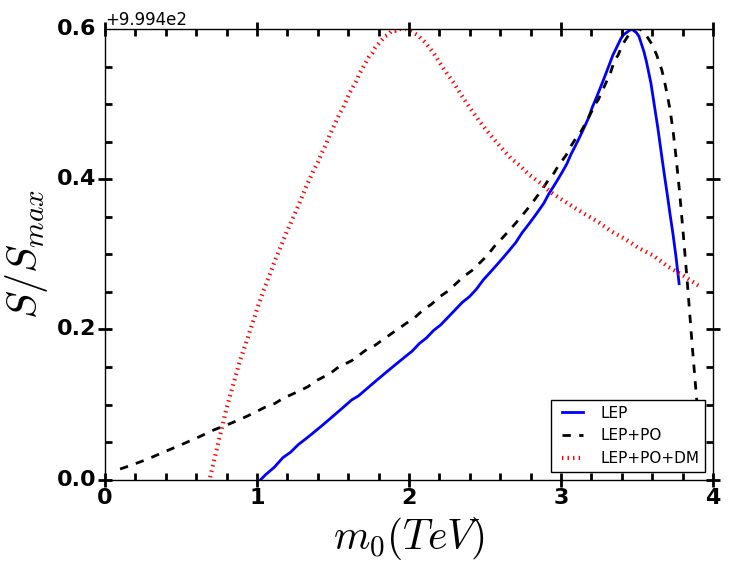}
\hspace{-0.9em}
\includegraphics[angle=0,width=0.327\linewidth,height=15em]{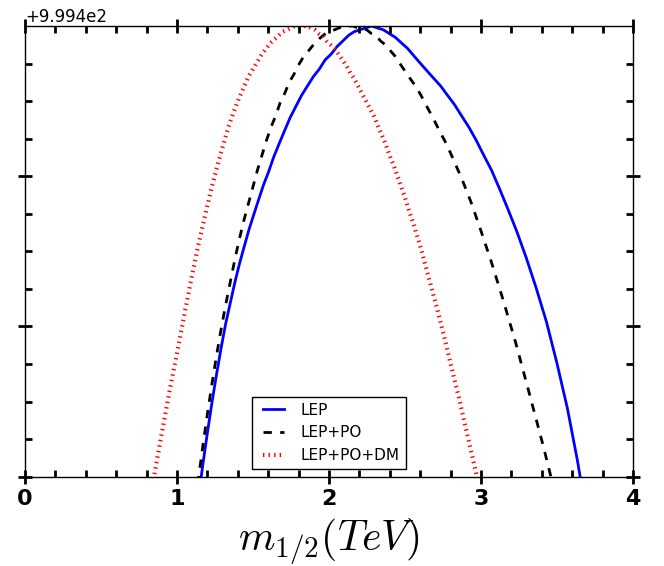}
\hspace{-1.0em}
\includegraphics[angle=0,width=0.327\linewidth,height=15em]{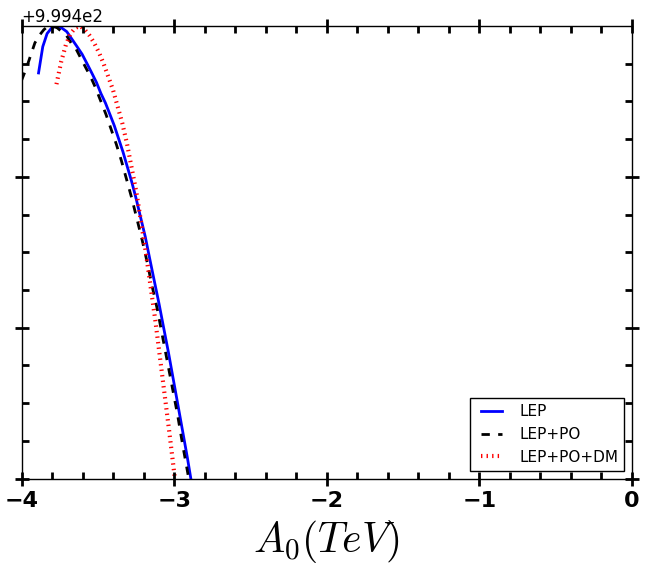}
\includegraphics[angle=0,width=0.327\linewidth,height=15.2em]{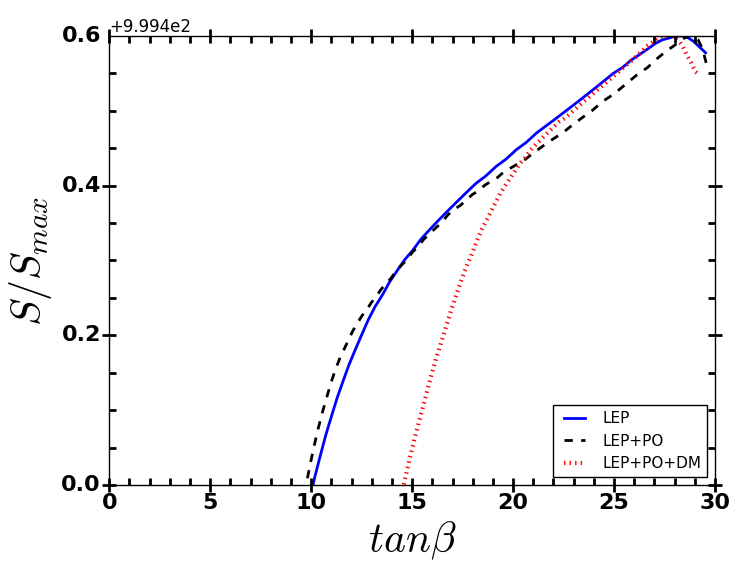}
\hspace{-0.9em}
\includegraphics[angle=0,width=0.327\linewidth,height=15em]{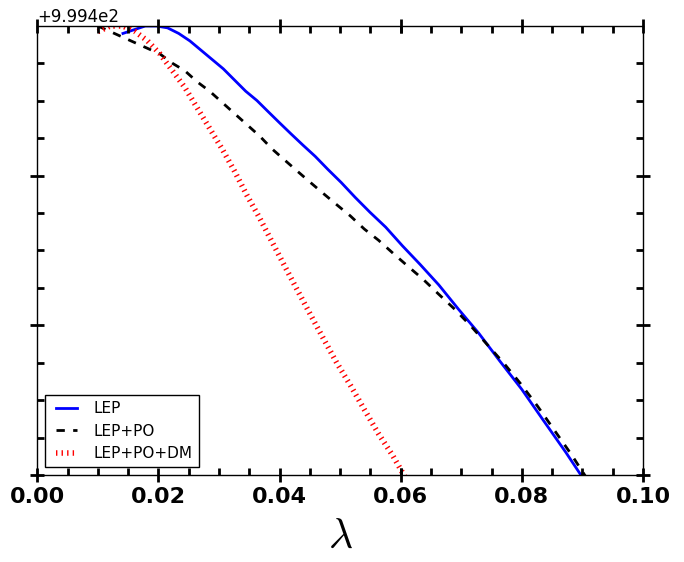}
\hspace{-0.9em}
\includegraphics[angle=0,width=0.327\linewidth,height=15em]{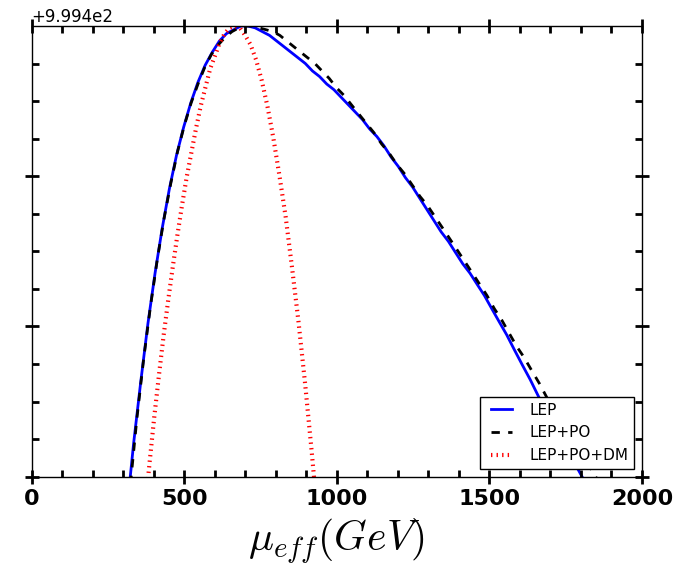}
\caption{\sf{Entropy vs parameters corresponding to various constraints. The color scheme is same as the Figure~\ref{fig1}.}}
\label{fig4}
\end{centering}
\end{figure}

Information entropy can be able to estimate using Eq.~\ref{3.2e} considering an ensemble of $\cal N$-independent Higgs bosons which allow decaying into its allowed decay modes, namely $h\rightarrow\gamma\gamma$, $h\rightarrow \gamma Z$, $h\rightarrow Z Z^*$, $h\rightarrow W W^*$, $h\rightarrow gg$, $h\rightarrow f\bar{f}$ where $f\in \{u, d, c, s, b, e, \mu, \tau \}$, $h\rightarrow A A$. Then Higgs entropy can be further used to estimate the mass of the parameters. 
The information entropy would be a function of the Higgs mass $m_h$ only marginalised on other parameters followed by scaling with a normalising factor $1/S_{max}$. The parameter space has been restricted with LEP data on masses of Higgs boson, neutralino and chargino such as $m_h >$ 114.4 GeV, $ m_{\tilde\chi^{0}_{1,2,3,4}} >$ 0.5 $m_Z$ and $ m_{\tilde\chi^{\pm}_{1,2}} >$ 103.5 GeV, respectively, branching ratio of B-Physics $ BR(B^0_s \to \mu^+\mu^-)$, and the relic density of the dark matter $\Omega_{\chi}h^2$ at 2.5$\sigma$ confidence level as mentioned in Table~\ref{tab:table1}. Our results described in Figures~\ref{fig1}--\ref{fig4} taking into consideration the constraints as follows (a) LEP, (b) LEP$+$PO, and (c) LEP$+$PO$+$DM. Here, the blue solid line shows the mass limit from LEP data on Higgs boson, neutralinos and charginos, the black dashed line signifies the constraint on LEP data and B-Physics, and the red dotted line includes the constraints on LEP data, B-Physics, and the relic density of neutralino dark matter.

We present the plots to illustrate the variation of information entropy with CP-even Higgs boson and CP-odd Higgs boson in Figure~\ref{fig1}, with neutralino, chargino, and gluino in Figure~\ref{fig2}, with stops, sbottoms, and staus in Figure~\ref{fig3}, and with free parameters in Figure~\ref{fig4}. It is to be noted that the SM-like Higgs is in good agreement with the LHC measured Higgs mass. The associated values of $m_0$, $m_{1/2}$, $ A_0$, $ tan\beta$, $\lambda$, $\mu_{eff}$, lightest CP-odd Higgs boson $m_{A_1}$, LSP neutralino, singlino, lightest chargino, gluino, lightest stop and lightest stau at maximum entropy against the LEP constraints are observed to be 3.46 TeV, 2.28 TeV, $-$3.77 TeV, 28.1, 0.018, 685.9 GeV, 95.2 GeV, 1.49 TeV, 11.35 TeV, 1.61 TeV, 4.48 TeV, 2.83 TeV, and 4.15 TeV, respectively, while the corresponding values are found to be 3.49 TeV, 2.10 TeV, $-$3.82 TeV, 28.6, 0.01, 701.3 GeV, 110.2 GeV, 1.48 TeV, 10.71 TeV, 1.74 TeV, 4.45 TeV, 2.86 TeV, and 4.14 TeV, respectively, in concern of LEP data and B-Physics branching ratio. The corresponding values for the above-described parameters are obtained to be 1.93 TeV, 1.78 TeV, $-$3.62 TeV, 27.5, 0.012, 665.7 GeV, 110.3 GeV, 0.74 TeV, 11.24 TeV, 0.79 TeV, 3.70 TeV, 0.83 TeV, and 1.19 TeV, respectively, after taking into account constraints of LEP data, B-Physics branching ratio, and the relic density of neutralino dark matter. 

It is worth noting that in our work, the CP-odd Higgs boson is observed as being the lightest among the Higgs bosons~\cite{Ellwanger:2008ya}. The mechanisms of dark matter constraints has been discussed in Ref.~\cite{Costa:2017gup}. The observed parameters values and the sparticles masses after taking into account the constraint of relic density of dark matter along with various constraints in our work preferred the stop, chargino, as well as focus point coannihilation regions satisfying their respective conditions given in Ref.~\cite{Costa:2017gup}. It is noteworthy that the LSP neutralino would observe a mass below a TeV. It is to be noted that the value of $\mu_{eff}$ is observed under a TeV, thereby solving the problem of $\mu$. The most preferred values of sparticles masses of the parameters are mentioned in Table~\ref{tab:table2}.

\section{Summary and Conclusions}

In the current study, we have explored the NMSSM scenario in the light of Higgs entropy built using its Higgs branching ratios corresponding to allowed decay modes taking into consideration of LEP data, B-Physics, and dark matter relic density. Moreover, Higgs entropy makes it possible to estimate the masses of the parameters. We have presented the variation of information entropy versus masses of Higgs, masses of neutralino and chargino, masses of squarks, and parameters in Figures~\ref{fig1}--\ref{fig4}. Considering the information-theoretic approach, in the concern of LEP data constraints, the corresponding values of $m_0$, $m_{1/2}$, $A_0$,  $ tan\beta$,  $\lambda$, and  $\mu_{eff}$ are found to be 3.46 TeV, 2.28 TeV, $-$3.77 TeV, 28.1, 0.018, 685.9 GeV while in the consideration of constraints from LEP data and B-physics branching ratios, the corresponding values are observed to be 3.49 TeV, 2.10 TeV, $-$3.82 TeV, 28.6, 0.01, 701.3 GeV. After taking into account LEP data, B-Physics, and dark matter relic density, the corresponding values of the above-mentioned parameters are found to be 1.93 TeV, 1.78 TeV, $-$3.62 TeV, 27.5, 0.012, 665.7 GeV. The corresponding mass values of SM-like Higgs boson $m_h$, lightest CP-odd Higgs boson $m_{A_1}$, mass neutralino LSP $ m_{\tilde\chi^{0}_{1}}$, lightest chargino $ m_{\tilde\chi^{\pm}_{1}}$, singlino $ m_{\tilde\chi^{0}_{5}}$, and gluino $ m_{\tilde g}$ are expected to be 125.2 GeV, 95.2 GeV, 1.49 TeV, 1.61 TeV, 11.35 TeV, and 4.48 TeV, respectively, under the constraints of LEP, 125.4 GeV, 110.2 GeV, 1.48 TeV, 1.74 TeV, 10.71 TeV, and 4.45 TeV, respectively, after taking into account the LEP and B-physics branching ratio, and 125.5 GeV, 110.3 GeV, 0.74 TeV, 0.79 TeV, 11.24 TeV, and 3.70 TeV, respectively, under the constraints of LEP, B-Physics and dark matter relic density. It is worth noting that the observed mass of the Higgs boson is in good accord with the experimentally measured value of the Higgs boson at the LHC ($m_h =125.25\pm0.17$)~\cite{Workman:2022ynf}. Our study indicates that the mass of neutralino LSP and $\mu_{eff}$ are observed to be 0.74 TeV and 0.69 TeV, respectively, (below a TeV) after taking into account various constraints and the relic density of dark matter. 

 
 \begin{table}[h!]
  \begin{center}
 \tabcolsep 0.8pt
    \small
    \begin{tabular}{ccccc} 
      \hline
      \hline
        \multirow{2}{*}{Parameter}& \multicolumn{3}{c}{Constraints}& \\
      \cline{2-4} 
      & {LEP} & {LEP $+$ PO}&\hspace{0.3em} {LEP $+$ PO $+$ DM} \\
       \hline
       \hline   
      $m_0$ &3.46&3.49&1.93&\\ 
      $ m_{1/2}$  &2.28&2.10&1.78&\\
      $ A_0$  &$-$ 3.77&$-$ 3.82&$-$3.62&\\
      $ tan\beta$  &28.1&28.6&27.5&\\
      $\lambda$  &0.018&0.01&0.012&\\
      $\mu_{eff}$ (GeV)  &685.9&701.3&665.7&\\
           \hline 
           \hline
      $m_h$ (GeV)&125.2&125.4&125.5&\\
      $m_{H_1}$&23.12&23.16&8.82&\\
      $m_{H_2}$&15.67&15.71&9.75&\\
      $m_{A_1}$ (GeV)&95.2&110.2&110.3&\\
      $m_{A_2}$&23.68&23.68&9.22&\\
      $m_{H_\pm}$&23.31&23.53&9.69&\\
      \hline       
      $ m_{\tilde\chi^{0}_{1}}$ &1.49&1.48&0.74&\\
      $ m_{\tilde\chi^{0}_{2}}$   &2.48&1.62&0.77&\\
       $ m_{\tilde\chi^{0}_{3}}$ &1.54&1.55&1.16&\\
      $ m_{\tilde\chi^{0}_{4}}$  &2.09&2.1&2.27&\\
      $ m_{\tilde\chi^{0}_{5}}$  &11.35&10.71&11.24&\\
      $ m_{\tilde\chi^{\pm}_{1}}$  &2.33&2.36&0.79&\\
      $ m_{\tilde\chi^{\pm}_{2}}$  &1.61&1.74&1.57&\\
      $ m_{\tilde g}$ &4.48&4.45&3.70&\\ 
         \hline  
      $ m_{\tilde \tau_1}$ &4.15&4.14&1.93&\\ 
      $ m_{\tilde \tau_2}$ &6.17&6.16&1.19&\\     
      $ m_{\tilde t_1}$ &2.83&2.86&0.83&\\
      $ m_{\tilde t_2}$ &4.07&4.09&3.23&\\
      $ m_{\tilde b_1}$ &4.09&4.08&3.19&\\
      $ m_{\tilde b_2}$ &4.61&4.55&3.6&\\
                 \hline
                 \hline
    \end{tabular}
    \caption{\sf{Allowed NMSSM parameters and its respective spectrum corresponding to maximum entropy. The masses are in TeV unless otherwise specified.}}
       \label{tab:table2}
 \end{center}
\end{table}


\section*{Acknowledgement}
The present work has been supported in part by the University Grant Commission through Start-up Grant No. F30-377/2017 (BSR). We are grateful to Apurba Tiwari for her helpful discussions.
We acknowledge the DST Computational laboratory facility available at the Department of Physics, AMU, Aligarh, India during the initial phase of the work.



\begin{thebibliography}{44}



\bibitem{ATLAS:2012yve}
G.~Aad \textit{et al.} [ATLAS],
Phys. Lett. B \textbf{716}, 1-29 (2012)
doi:10.1016/j.physletb.2012.08.020
[arXiv:1207.7214 [hep-ex]].


\bibitem{CMS:2012qbp}
S.~Chatrchyan \textit{et al.} [CMS],
Phys. Lett. B \textbf{716}, 30-61 (2012)
doi:10.1016/j.physletb.2012.08.021
[arXiv:1207.7235 [hep-ex]].



\bibitem{Aad:2015zhl}
G.~Aad {\it et al.} [ATLAS and CMS],
Phys.\ Rev.\ Lett.\ {\bf 114}, 191803 (2015),
[arXiv:1503.07589 [hep-ex]].



\bibitem{Djouadi:2005gi}
A.~Djouadi,
Phys. Rept. \textbf{457}, 1-216 (2008)
doi:10.1016/j.physrep.2007.10.004
[arXiv:hep-ph/0503172 [hep-ph]].


\bibitem{Martin:1997ns}
S.~P.~Martin,
Adv.\ Ser.\ Direct.\ High Energy Phys.\ {\bf 21}, 1-153 (2010),
[arXiv:hep-ph/9709356 [hep-ph]].


\bibitem{Tata:1997uf}
X.~Tata,
[arXiv:hep-ph/9706307 [hep-ph]].


\bibitem{Drees:1996ca}
M.~Drees,
[arXiv:hep-ph/9611409 [hep-ph]].


\bibitem{Fayet:2015sra}
P.~Fayet,
Adv.\ Ser.\ Direct.\ High Energy Phys.\ {\bf 26}, 397-454 (2016),
[arXiv:1506.08277 [hep-ph]].


\bibitem{Cane:2019ac}
A.~Canepa,
Rev.\ Phys.\ {\bf 4}, 100033 (2019).


\bibitem{Allanchach:2019wrx}
B.~C.~Allanchach,
CERN Yellow Rep.\ School Proc.\ {\bf 6}, 113-144 (2019).


\bibitem{Aitchison:2005cf}
I.~J.~R.~Aitchison,
[arXiv:hep-ph/0505105 [hep-ph]].


\bibitem{Djouadi:2005} 
A.~Djouadi,
Phys.\ Rept.\ {\bf 459}, 1-241 (2008),
[arXiv:hep-ph/0503173 [hep-ph]].


\bibitem{Nilles:1983ge}
H.~P.~Nilles,
Phys. Rept. \textbf{110}, 1-162 (1984)
doi:10.1016/0370-1573(84)90008-5


\bibitem{Haber:1984rc}
H.~E.~Haber and G.~L.~Kane,
Phys. Rept. \textbf{117}, 75-263 (1985)
doi:10.1016/0370-1573(85)90051-1

\bibitem{Dawson:1997tz}
S.~Dawson,
[arXiv:hep-ph/9712464 [hep-ph]].


\bibitem{Heinemeyer:1998np}
S.~Heinemeyer, W.~Hollik and G.~Weiglein,
Eur.\ Phys.\ J.\ C {\bf 9}, 343-366 (1999),
[arXiv:hep-ph/9812472 [hep-ph]].


\bibitem{Draper:2016pys}
P.~Draper and H.~Rzehak,
Phys.\ Rept.\ {\bf 619}, 1-24 (2016),
[arXiv:1601.01890 [hep-ph]].



\bibitem{Ellwanger:2009dp}
U.~Ellwanger, C.~Hugonie and A.~M.~Teixeira,
Phys.\ Rept.\ {\bf 496}, 1-77 (2010),
[arXiv:0910.1785 [hep-ph]].


\bibitem{Maniatis:2009re}
M.~Maniatis,
Int. J. Mod. Phys. A \textbf{25}, 3505-3602 (2010)
[arXiv:0906.0777 [hep-ph]].



\bibitem{Kim:1983dt}
J.~E.~Kim and H.~P.~Nilles,
Phys. Lett. B \textbf{138}, 150-154 (1984)


\bibitem{Ellwanger:2005uu}
U.~Ellwanger, J.~F.~Gunion and C.~Hugonie,
JHEP \textbf{07}, 041 (2005)
doi:10.1088/1126-6708/2005/07/041
[arXiv:hep-ph/0503203 [hep-ph]].


\bibitem{Ellwanger:2011sk}
U.~Ellwanger,
Eur. Phys. J. C \textbf{71}, 1782 (2011)
[arXiv:1108.0157 [hep-ph]].


\bibitem{Heng:2023xfb}
Z.~Heng, S.~Yang, X.~Li and L.~Shang,
Symmetry {\bf 15}, no.2, 456 (2023)
[arXiv:2302.07465 [hep-ph]].


\bibitem{Beskidt:2017dil}
C.~Beskidt, W.~de Boer and D.~I.~Kazakov,
Phys. Lett. B \textbf{782}, 69-76 (2018)
[arXiv:1712.02531 [hep-ph]].


\bibitem{Barman:2020vzm}
R.~K.~Barman, G.~B\'elanger, B.~Bhattacherjee, R.~Godbole, D.~Sengupta and X.~Tata,
Phys. Rev. D \textbf{103}, no.1, 015029 (2021)
[arXiv:2006.07854 [hep-ph]].


\bibitem{Biekotter:2021qbc}
T.~Biek\"otter, A.~Grohsjean, S.~Heinemeyer, C.~Schwanenberger and G.~Weiglein,
Eur. Phys. J. C \textbf{82}, no.2, 178 (2022)
[arXiv:2109.01128 [hep-ph]].


\bibitem{Tang:2022pxh}
T.~P.~Tang, M.~Abdughani, L.~Feng, Y.~L.~S.~Tsai, J.~Wu and Y.~Z.~Fan,
Sci. China Phys. Mech. Astron. \textbf{66}, no.3, 239512 (2023)
[arXiv:2204.04356 [hep-ph]].


\bibitem{dEnterria:2012eip}
D.~d'Enterria,
``On the Gaussian peak of the product of decay probabilities of the standard model Higgs boson at a mass $m_H\sim$125 GeV,''
[arXiv:1208.1993 [hep-ph]].



\bibitem{Alves:2014ksa}
A.~Alves, A.~G.~Dias and R.~da Silva,
Physica A {\bf 420}, 1-7 (2015),
[arXiv:1408.0827 [hep-ph]].


\bibitem{Alves:2020cmr}
A.~Alves, A.~G.~Dias and R.~da Silva,
Nucl.\ Phys.\ B {\bf 959}, 115137 (2020),
[arXiv:2004.08407 [hep-ph]].


\bibitem{Llanes-Estrada:2017clj}
F.~J.~Llanes-Estrada, P.~C.~Millan, A.~Porras Riojano, E.~M.~Sánchez García and M.~Á.~García Ferrero,
PoS {\bf EPS-HEP2017}, 740 (2017),
[arXiv:1710.01286 [hep-ph]].


\bibitem{Millan:2018fme}
P.~Carrasco Millán, M.~Á.~García-Ferrero, F.~J.~Llanes-Estrada, A.~Porras Riojano and E.~M.~Sánchez García,
Nucl.\ Phys.\ B {\bf 930}, 583-596 (2018),
[arXiv:1802.05487 [hep-ph]].


\bibitem{Alves:2017ljt}
A.~Alves, A.~G.~Dias and R.~Silva,
Braz.\ J.\ Phys.\ {\bf 47}, no.4, 426-435 (2017),
[arXiv:1703.02061 [hep-ph]].


\bibitem{Gupta:2020whs}
S.~Gupta and S.~Kumar Gupta,
Nucl.\ Phys.\ B {\bf 965}, 115336 (2021),
[arXiv:2008.00415 [hep-ph]].


\bibitem{Gupta:2022psc}
S.~Gupta and S.~K.~Gupta,
Nucl.\ Phys.\ B {\bf 984}, 115942 (2022),
[arXiv:2205.00173 [hep-ph]].


\bibitem{Gupta:2022mjt}
S.~Gupta and S.~K.~Gupta,
Nucl. Phys. B \textbf{986}, 116056 (2023)
[arXiv:2208.06627 [hep-ph]].



\bibitem{jaynes:1957}
E.~T.~Jaynes, 
Phys.\ Rev.\ {\bf 106}, 620-630 (1957).

 
\bibitem{thomas:2006}
T.~M.~Cover and J.~A.~Thomas, 
Second Edition, Wiley-Interscience (2006). 


\bibitem{shannon}
C.~E.~Shannon, 
The Bell Syst.\ Tech.\ J.\ {\bf 27}, 379-423 (1948).



\bibitem{Workman:2022ynf}
R.~L.~Workman \textit{et al.} [Particle Data Group],
PTEP \textbf{2022}, 083C01 (2022),


\bibitem{ALEPH:2006tnd}
S.~Schael \textit{et al.} [ALEPH, DELPHI, L3, OPAL and LEP Working Group for Higgs Boson Searches],
Eur. Phys. J. C \textbf{47}, 547-587 (2006)
doi:10.1140/epjc/s2006-02569-7
[arXiv:hep-ex/0602042 [hep-ex]].


\bibitem{Zyla:2020zbs}
P.~A.~Zyla {\it et al.} [Particle Data Group], 
Prog.\ Theor.\ Exp.\ Phys.\ {\bf 2020}, no.8, 083C01 (2020).


\bibitem{Planck:2015fie}
P.~A.~R.~Ade \textit{et al.} [Planck],
Astron.\ Astrophys.\ {\bf 594}, A13 (2016),
[arXiv:1502.01589 [astro-ph.CO]].


\bibitem{Belanger:2013oya}
G.~Belanger, F.~Boudjema, A.~Pukhov and A.~Semenov,
Comput. Phys. Commun. \textbf{185}, 960-985 (2014)
doi:10.1016/j.cpc.2013.10.016
[arXiv:1305.0237 [hep-ph]].


\bibitem{Ellwanger:2004xm}
U.~Ellwanger, J.~F.~Gunion and C.~Hugonie,
JHEP \textbf{02}, 066 (2005)
[arXiv:hep-ph/0406215 [hep-ph]].



\bibitem{Ellwanger:2005dv}
U.~Ellwanger and C.~Hugonie,
Comput. Phys. Commun. \textbf{175}, 290-303 (2006)
doi:10.1016/j.cpc.2006.04.004
[arXiv:hep-ph/0508022 [hep-ph]].



\bibitem{Ellwanger:2008ya}
U.~Ellwanger,
AIP Conf.\ Proc.\ {\bf 1078}, no.1, 73-78 (2009)
[arXiv:0809.0779 [hep-ph]]


\bibitem{Costa:2017gup}
J.~C.~Costa, E.~Bagnaschi, K.~Sakurai, M.~Borsato, O.~Buchmueller, M.~Citron, A.~De Roeck, M.~J.~Dolan, J.~R.~Ellis and H.~Fl\"acher, \textit{et al.}
Eur.\ Phys.\ J.\ C {\bf 78}, no.2, 158 (2018),
[arXiv:1711.00458 [hep-ph]].


\end{thebibliography}
\end{document}